\documentclass[twocolumn,prl,aps,floatfix,superscriptaddress,amssymb]{revtex4}
\usepackage{epsfig}

\begin{document}

\title{Magnetism and correlations in fractionally filled degenerate shells of \\
  graphene quantum dots}

\author{A.~D.~G\"u\c{c}l\"u}
\affiliation{Institute for Microstructural Sciences, National Research Council of Canada
, Ottawa, Canada}

\author{P.~Potasz}
\affiliation{Institute for Microstructural Sciences, National Research Council of Canada
, Ottawa, Canada}
\affiliation{Institute of Physics, Wroclaw University of Technology, Wroclaw, Poland}

\author{O.~Voznyy}
\affiliation{Institute for Microstructural Sciences, National Research Council of Canada
, Ottawa, Canada}

\author{M.~Korkusinski}
\affiliation{Institute for Microstructural Sciences, National Research Council of Canada
, Ottawa, Canada}

\author{P.~Hawrylak}
\affiliation{Institute for Microstructural Sciences, National Research Council of Canada
, Ottawa, Canada}

\date{\today}

\begin{abstract}
We investigate interaction effects and magnetization in triangular graphene
quantum dots using a combination of tight-binding, Hartree-Fock and
configuration interaction methods. We show that electronic correlations play a
crucial role in determining the nature of the ground state as a function of
filling fraction of the degenerate shell at the Fermi level. A
half-filled  charge neutral shell leads to full spin polarization but this
magnetic moment can be completely destroyed by adding a single electron.
\end{abstract}

\maketitle

Following the progress in the fabrication of
graphene\cite{NGM+04,NGM+05,ZTS+05,ZGG+06,NGP+09} based devices, lower
dimensional structures such as graphene
ribbons\cite{NFD+96,WFA+99,WSS+08,YCL08}, and more recently graphene  quantum
dots\cite{WSG08,WRA+09,AB08,LSB09,ZCP08} are attracting increasing attention
due to their non-trivial electronic and magnetic properties. In particular, it
was shown that when an electron is confined to a triangular atomic thick layer
of graphene with zig-zag edges, its energy spectrum collapses to a shell of
degenerate states at the Fermi level  (Dirac
point)\cite{Eza07,Eza08,FP07,WMK08,AHM08} isolated from remaining states by a
gap. The degeneracy is proportional to the edge size and can be made
macroscopic. This opens up the possibility to design a strongly correlated
electronic system as a function of fractional filling of the shell, in analogy
to the fractional quantum Hall effect\cite{TSG82}, but without the need for a
magnetic field. 

In this work, we present new results demonstrating the important role of
electronic correlations, beyond the Hubbard model\cite{Eza07,Eza08,FP07} and
mean-field density functional theory (DFT)\cite{FP07,WMK08}.  The interactions
are treated by a combination of DFT, tight-binding, Hartree-Fock and
configuration interaction methods (TB-HF-CI).
We show that a half-filled charge neutral shell leads to full
spin polarization of the island but this magnetic moment is completely
destroyed by the addition of a single electron, in analogy to the effect of
skyrmions on the quantum Hall  ferromagnet\cite{SKK+93,MFB96,NM06,PSM+09} and
spin depolarization in electrostatically defined semiconductor quantum
dots\cite{Haw93,PMC+94,KHC+04,CSH+00}.  The depolarization of the ground state
is predicted to result in blocking of current through a graphene quantum dot
due to spin blockade (SB)\cite{PMC+94,CSH+00}.


In order to describe the interaction effects of electrons occupying the edge
states (see Fig.1), we first solve the mean-field problem using a combination of
tight-binding approach with a self-consistent Hartree-Fock method for the
system with empty edge states, i.e. with $N_{ref} = N_{site}-N_{edge}$
electrons, where $N_{site}$ is the number of atoms, and $N_{edge}$ is the number of edge
states. After extensive algebra, we arrive at the HF Hamiltonian given by
\begin{eqnarray}
\nonumber
&H_{MF}&= \sum_{i,l,\sigma}\tau_{il\sigma}c^\dagger_{i\sigma}c_{l\sigma} 
     + \sum_{i,l,\sigma} \sum_{j,k,\sigma'}(\rho_{jk}-\rho^{bulk}_{jk}) 
(\langle ij\vert V \vert kl\rangle \\
&-&\langle ij\vert V \vert lk\rangle\delta_{\sigma,\sigma'})
           c^\dagger_{i\sigma}c_{l\sigma} 
+ \sum_{i,\sigma}v^{g}_{ii}(q_{ind})c^\dagger_{i\sigma}c_{i\sigma} 
\label{hfhamilton}
\end{eqnarray}
where the operator $c^\dagger_{i\sigma}$  creates a $p_z$ electron on site
``i'' with spin $\sigma$. The tight-binding parameters $\tau_{il\sigma}$ are 
fixed to their bulk values -2.5eV for
nearest neighbours, and -0.1eV for next nearest neighbours. Within the
Hartree-Fock approximation, electrons interact with each other via the density
matrix $\rho=\rho^{\uparrow}+\rho^{\downarrow}$ from which we subtract the graphene bulk density matrix 
$\rho^{bulk}$ already present in the tight-binding term $\tau_{il\sigma}$. In addition to the on-site
interaction term, all scattering and exchange terms within next nearest
neighbours, and all direct interaction terms are included in the two-body
Coulomb matrix elements $\langle ij|V|kl\rangle$ computed using Slater $p_z$ orbitals. In
order to first empty the degenerate shell and later charge it with successive
electrons, we transfer electrons to a metallic gate. The electrons in the
graphene island interact with the gate via the term $v_{ii}^{g}$  given by
\begin{eqnarray}
v_{ii}^{g}(q_{ind})&=& \sum^{N_{site}}_{j=1}\frac{-q_{ind}/N_{site}}
              {\kappa\sqrt{(x_i-x_j)^2+(y_i-y_j)^2+d^2_{gate}}}
\label{gate}
\end{eqnarray}
where $(x_i,y_i)$ are the coordinates of the atoms. This model assumes that the
induced charge $q_{ind}=-N_{edge}$ is smeared out at positions $(x_i,y_i)$ at a distance
$d_{gate}$ from the quantum dot.

After diagonalizing the Hamiltonian,Eq.(1), we
obtain TB+HF quasi-particles denoted by the creation operator $b^\dagger_ p$ , with
eigenvalues $\epsilon_p$ and eigenfunctions $|p>$  shown in Fig.1. We then start filling
the edge states one by one, and solve the configuration interaction problem
for the $N_{add}$ added electrons using 
\begin{eqnarray}
\nonumber
H&=& \sum_{p,\sigma}\epsilon_{p}b^\dagger_{p\sigma}b_{p\sigma}
    +\frac{1}{2}\sum_{p,q,r,s,\sigma \sigma'}\langle pq\vert V \vert rs\rangle 
     b^\dagger_{p\sigma}b^\dagger_{q\sigma'}b_{r\sigma'}b_{s\sigma} \\
\nonumber
  &+& \sum_{p,q,\sigma} \langle p\vert v^{g}(N_{add})\vert q\rangle
     b^\dagger_{p\sigma}b_{q\sigma}  \\
  &+& 2\sum^{N_{ref}/2}_{p'} \langle p' \vert v^{g}(N_{add})\vert p'\rangle
\label{CIhamilton}
\end{eqnarray}
where the indices $(p,q,r,s)$ run over edge states, while the index $p'$ runs over
valence states (below the edge states).




\begin{figure}
\epsfig{file=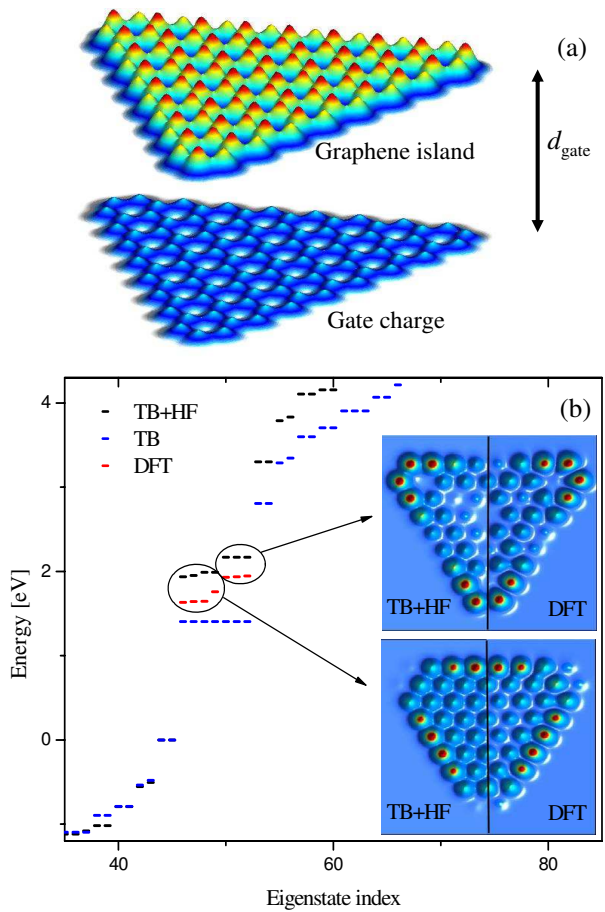,width=3.2in}
\caption{\label{fig:meanfield}
(Color online) 
(a) Electronic density in a triangular graphene island of 97 carbon
atoms where 7 electrons were moved to the metallic gate at a distance of
$d_{gate}$. (b)  Single particle spectrum of the structure in (a), obtained by
tight-binding (TB, blue lines) and self-consistent Hartree-Fock (TB+HF, black
lines)  methods. The 7 edge states near the Fermi level are compared  to DFT
results. In Hartree-Fock and DFT calculations 7 electrons were removed,
leaving the zero-energy states empty. The dielectric constant $\kappa$ is set to
6. Inset compares the structure of corner and side states obtained using
Hartree-Fock and DFT calculations. In DFT calculations, hydrogen atoms were
attached to dangling bonds.}
\end{figure}
Fig.1a shows the electronic density of a zig-zag edged triangular island of
$N=97$ carbon atoms, separated by a distance $d_{gate}$ from a metallic
gate. At zero applied voltage the island is charge neutral while applied
voltage leads to removal/addition of electrons to the island. The
single-particle energy spectrum obtained using nearest neighbour tight-binding
method (TB, blue lines) and a combination of the next-nearest neighbour
tight-binding and Hartree-Fock methods (TB+HF, black lines) is shown in
Fig.1b. As was previously shown by Ezawa\cite{Eza08}, Fernandez-Rossier and
Palacios\cite{FP07} and Wang, Meng and Kaxiras\cite{WMK08} using
nearest-neighbour TB method and ab-initio DFT calculations, the linear
spectrum of Dirac electrons in bulk graphene collapses to a shell of
degenerate levels at the Fermi energy, well separated in energy from the
valence and conduction bands. Similar to edge states in graphene
ribbons\cite{NFD+96,WFA+99,WSS+08,YCL08}, zigzag edge breaks the symmetry
between the two sublattices of the honeycomb lattice, behaving like a
defect. Therefore, electronic states localized on the zigzag edges appear with
energy in the vicinity of the Fermi level. For a $N=97$ atoms island (see
Fig.1) there are $N_{edge}=7$ edge states. For the charge neutral system there
is one electron per each edge state. A non-trivial question addressed here is
the specific spin and orbital configuration of the electrons as a function of
the size and the fractional filling of the degenerate shell of edge
states. Due to the strong degeneracy, many-body effects can be expected to be
important as in the fractional quantum Hall effect. Previous calculations
based on the Hubbard approximation\cite{Eza07,Eza08,FP07} and local spin
density functional  theory\cite{FP07,WMK08} showed that the  neutral system
(half-filling) has its edge states polarized. 

In order to study many-body effects within the charged  shell of the edge
states via the configuration interaction method, we first perform a
Hartree-Fock calculation for the charged system of $N-N_{edge}$ electrons,
with empty edge states and $N_{edge}$ electrons transferred to the gate, shown
in Fig.1a.  The spectrum of HF quasi-particles is shown with black lines in
Fig.1b. Due to the mean-field interaction with the valence electrons and
charged gate, the degeneracy of the edge states is lifted, and the
electron-hole symmetry is broken. The main effect is that a group of three
states is now separated from the rest by a gap of $\sim 0.2$ eV. The three states
correspond to HF quasiparticles localised in the three corners of the
triangle.   The same physics occurs in density functional calculation within
local density approximation (LDA), shown with red lines in Fig.1b. Hence we
see that the shell of edge states with a well defined gap separating them from
the valence and conduction bands exists in the three approaches. 

\begin{figure}
\epsfig{file=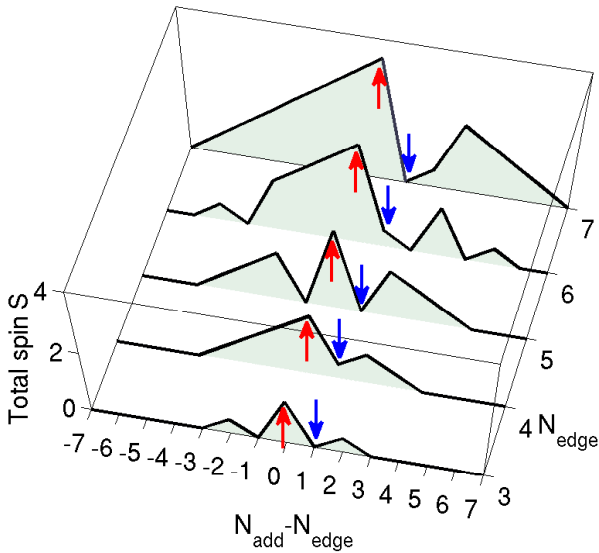,width=3.2in}
\caption{\label{fig:spinphase}
(Color online) 
Spin phase diagram from TB-HF-CI method as a function of the size of
the triangular dot characterized by the number of edge states $N_{edge}$, and the
filling of the zero-energy states Nadd-Nedge. Charge neutral case corresponds
to $N_{add}-N_{edge}=0$, for which the total spin of the zero-energy electrons are
always maximized ($S=N_{edge}/2$, indicated by red arrows). On the other hand, if the
quantum  dot is charged by 1 electron ($N_{add}-N_{edge}=1$) then the total spin has
minimum value, i.e. $S=0$ if $N_{add}$ is even, $S=1/2$ if $N_{add}$ is odd (indicated by
blue arrows).}
\end{figure}
The wave functions corresponding to the band of nearly-degenerate edge states
obtained from TB-HF calculations are used as a basis set in our configuration
interaction calculations where we add $N_{add}$ electrons from the gate to the
shell of degenerate states. In Fig.2, total spin $S$ of the ground state as a
function of the filling number of the edge states is shown for different sizes
of quantum dots. Two aspects of these results are particularly
interesting. For the charge neutral case ($N_{add}-N_{edge}=0$),for all the island
sizes studied ($N_{edge}=3-7$), the half-filled shell is maximally spin polarized
as indicated by red arrows. This is in agreement with our DFT calculations and
reproduces previous work\cite{FP07,WMK08}. However, the spin polarisation is extremely
fragile. If we add one extra electron ($N_{add}-N_{edge}=1$), magnetisation of the
island collapses to the minimum possible value, as indicated by blue arrows in
Fig.2. We note that full or partial depolarization occurs for other filling
numbers but we focus here on the spin depolarisation at half-filling.  The
spin depolarisation was found to be insensitive to the screening of
electron-electron interactions, as values of dielectric screening constant $\kappa$
between 2 and 8 led to the same behaviour of the charge neutral and single
electron charged cases (not shown).

\begin{figure}
\epsfig{file=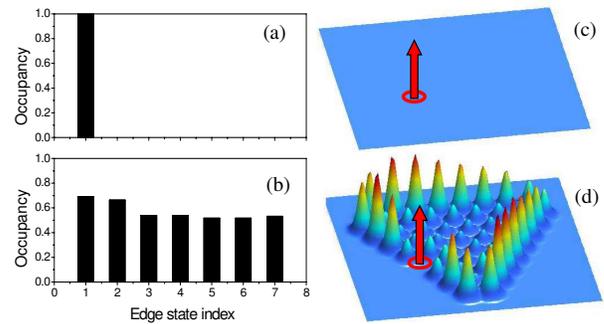,width=3.2in}
\caption{\label{fig:occup}
(Color online) 
Left panel: Orbital occupancy of the 7 edge states by spin up
electrons, for the charged ($N_{add}-N_{edge}=1$) system, for (a) $S=6/2$ and (b) $S=0$
total spin states. The ground state is $S=0$ (see Fig.2). Right panel:
Corresponding spin up-up pair-correlation functions 
$\langle\rho_{\uparrow}({\bf r}_0)\rho_{\downarrow}({\bf r})\rangle$. The fixed
spin-up electron is represented by a red arrow, and its position r0 by a red
circle.
}
\end{figure}
In order to illuminate the
depolarization process as an electron is added to the charge neutral maximally
spin polarized system, in Fig.3a,b we show the orbital occupancy of up-spin
edge states at $N_{add}-N_{edge}=1$, for the fully polarized state $S=6/2$ (upper panel)
and for the ground state, $S=0$, (lower panel) for the $N_{site}=97$ atoms quantum
dot with 7 edge states shown in Fig.1. For the large spin $S=6/2$ case, the
added spin up electron simply occupies the orbital 1 and its spin is opposite
to the spins of other 7 electrons.  However, the true ground state has $S=0$,
with spin occupancy shown in the lower panel. The added electron causes
electrons already present to partially flip their spin, with spin up density
being delocalized over all the 7 orbitals in analogy to skyrmion-like
excitations in quantum dots and quantum Hall 
ferromagnets\cite{SKK+93,MFB96,NM06,PSM+09}. The
correlated nature of the $S=0$ spin depolarised ground state is illustrated by
the up-up spin pair correlation functions given by
$\langle\rho_{\uparrow}({\bf r}_0)\rho_{\downarrow}({\bf r})\rangle$, shown in
Fig.3c,d. The location of the fixed up-spin electron at site ${\bf r}_0$ is
schematically shown with an up arrow. The up-up spin correlation function for
$S=6/2$ spin polarized system is strictly zero as there are no other spin up
electrons. The spin correlation function for the spin depolarized ground state
with $S=0$ shows exchange hole at ${\bf r}_0$ , which extends to the nearest neighbours,
and, more interestingly, for larger $\vert {\bf r}_0-{\bf r}\vert$ , spin pair correlation function
reveal a spin texture: Beyond the exchange hole there is the formation of an
electronic cloud with positive magnetization which decreases and changes sign
at even larger distance, again consistent with the skyrmion 
picture\cite{SKK+93,MFB96}.

\begin{figure}
\epsfig{file=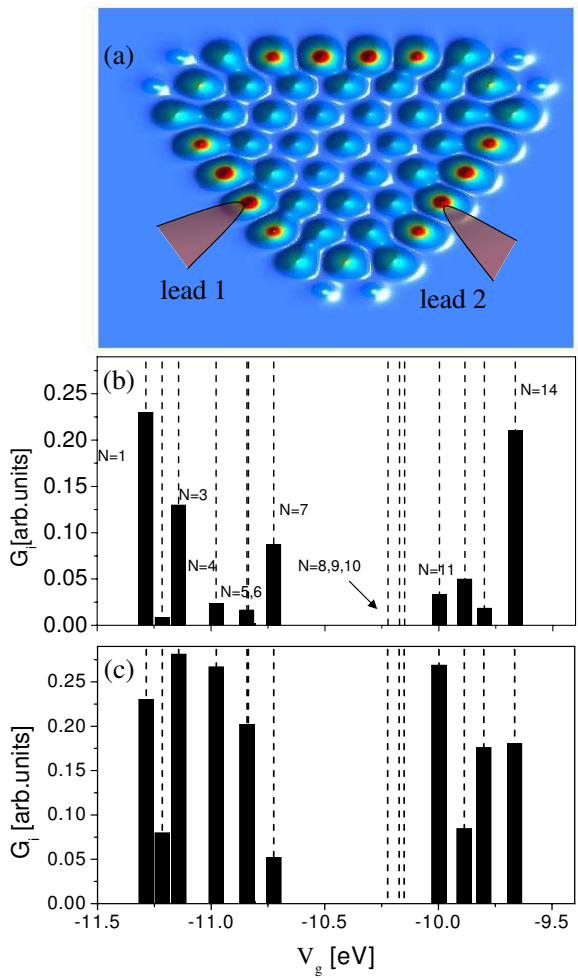,width=3.2in}
\caption{\label{fig:transport}
(Color online) 
(a) Schematic representation of the graphene island connected to the
leads through a side site. (b) Conductivity as a function of applied gate
voltage $V_g$ due to an electron entering the dot through the side atom. (c) Same
as (b) but without the site dependence of the incoming electron. The
oscillations of the spectral weight in (c) are purely to due correlation
effects and spin blockade.
}
\end{figure}
Experimentally, spin properties of quantum dots can be probed using Coulomb
and spin blockade spectroscopy\cite{CSH+00}. By connecting graphene quantum dot to
leads and measuring the conductance as a function of gate voltage, one obtains
a series of Coulomb blockade peaks. The relative position of these peaks and
their height reveal information about the electronic properties of the system
as the number of electrons is increased. The amplitude of the Coulomb blockade
peak is given by the conductivity $G_i$ of the graphene quantum dot connected to
leads via atom ``$i$''\cite{GSG+02} as shown schematically
in Fig.4a. Spin and correlation effects are reflected in the weight of Coulomb
blockade peak proportional to the matrix element 
$\vert \langle N+1,J',S' \vert c^\dagger_{i\sigma} \vert N,J,S \rangle\vert^2$
which gives the transition probability from state $(N,J,S)$ to state
$(N+1,J',S')$ when additional electron is added to the site ``$i$'' of
the graphene quantum dot from the lead.  The ground state configuration
$(N,J,S)$ is controlled by the gate voltage. For our model graphene quantum dot
with $N=7$ degenerate zero energy states, we can add a total of 14
electrons. Hence, one expects to obtain 14 peaks. In Fig.4b some of the peaks
have zero height due to spin blockade phenomenon. For instance, the transitions
from $(N=7,S=7/2)$ states to $(N=8,S=0)$ states are spin blocked since it is not
possible to change the spin of the system by $\Delta S=-7/2$ by adding one electron
with $S=1/2$. Similarly, transitions from $(N=9,S=1/2)$ states to $(N=10,S=4/2)$
states are spin blocked. Besides the spin blockade, one sees a strongly
oscillating feature of the spectral function heights. This is due to (a)
strongly correlated nature of the states $\vert N,S\rangle$ , and (b) specific choice of
the site ``$i$'', where the lead is attached to. Here, we chose a site close
to the middle of one of the sides of the triangle. The overlap of the site
wavefunction is strongly dependent on the nature of the edge-states. In
particular, the existence of corner states, as discussed in Fig.1, strongly
affects the transition probabilities. To isolate the effect of correlation to
the lead's position, in Fig.4c we plot the conductivity assuming that
the weight of the site ``$i$'' is a constant independent of the
edge-state. As a result, the weights of spectral peaks are different, except
for $N=8,9,10$ where the spin blockade occurs. These results show how to detect
the spin depolarization in transport experiments. Ultimately, we show here
that one can design a strongly correlated electron system in carbon based
material whose magnetic properties can be controlled by applied gate
voltage.


{\it Acknowledgement}. The authors thank NRC-CNRS CRP, Canadian Institute for
Advanced Research, Institute for Microstructural Sciences, and QuantumWorks
for support.

\vspace*{-0.22in}


\end{document}